# THE LABORATORY OF MATERIAL ANALYSIS WITH ION BEAMS LAMFI - USP

Manfredo H. Tabacniks

Department of Applied Physics, Institute of Physics, University of São Paulo
Caixa Postal 66318, São Paulo, 05315-970, SP, Brazil.
e-mail: tabacniks@if.usp.br

## ABSTRACT

LAMFI is a laboratory dedicated to the development and application of ion beam techniques for the analysis of bulk materials and thin films. Its main facilities comprise an 1.7MV Pelletron tandem accelerator and two analytical setups, one mainly for Rutherford Backscattering Spectrometry (RBS), and channeling, and another, for Particle Induced X-ray Emission (PIXE) analysis. Since its beginning in March 1992, over 10 institutions involving about 68 researchers, 42 graduate and 9 undergraduate students have been direct or indirectly users at LAMFI.

## 1. INTRODUCTION

LAMFI (Laboratório de Análise de Materiais por Feixe Iônico) is a laboratory dedicated to the development and application of ion beam techniques for the analysis of bulk materials and thin films. LAMFI is the result of a joint effort of many research groups from the University of São Paulo, which, since the seventies, have been applying nuclear techniques for the analysis of air pollution, environmental samples, semiconductors and thin films. Its main facilities comprises an 1.7MV Pelletron tandem accelerator and two analytical setups, one mainly for Rutherford Backscattering Spectrometry (RBS), and channeling, and the other, for Particle Induced X-ray Emission (PIXE) analysis. LAMFI is a multi-user laboratory whose operation is supervised by a scientific committee formed by five representatives from the Institute of Physics (IFUSP) and two from the Engineering Faculty (EPUSP), from the University of São Paulo. LAMFI is operated and maintained by a technical staff whose coordinator is indicated by the scientific committee.

## 2. THE FACILITIES

Right after arrival of the accelerator in 1992, the LAMFI was first installed on the 9th floor of the Pelletron Accelerator building. This allowed its immediate operation, while waiting for the conclusion of the its own building. Though in temporary installations, all the necessary resources needed to fully operate the accelerator and the analytical setups were available. In November 96, LAMFI was disassembled and moved to a new building. At the same time, the former technical coordinator, Prof. Juan Carlos Acquadro, decided to retire being substituted by Prof. Manfredo H. Tabacniks. In September 1997, LAMFI resumed its operation in the new installations.

### 2.1. The Accelerator

The main equipment of LAMFI is a 1.7MV NEC 5SDH tandem Pelletron accelerator with a nitrogen gas stripper. Two ion sources, an RF Alphatross with a Rb charge exchange cell, for He beam, and a SNICS II for H, Li, C, O, Si and heavier beams; and a low energy switching magnet are the components of the beam injector assembly. The high energy switching magnet, which is also used to control the beam energy, has 4 output lines respectively at $\pm30°$ and $\pm15°$ plus the zero degree line. Two complete analytical setups, each one with its own high vacuum chamber, vacuum pumps, detectors and data acquisition electronics, are connected to the switching magnet. The setup connected to the -15° line is mainly used for RBS and FRS analysis, while the +30° line is connected to a home-made PIXE analysis setup optimized for environmental samples (Tabacniks, 1983).

### 2.2. The analytical setups

The RBS-FRS station, shown in Figure 1, is a NEC high vacuum chamber (43cm ID and 15cm high) with a computer controlled goniometer with 5 degrees of freedom (x, y, z, θ, and φ). Surface barrier detectors can be mounted at any angle on a turn table while a fixed SB detector at 170° is used as a monito. A third SB detector at 30° with a 7μm thick Al absorber foil is used for hidrogen analysis by FRS with a He beam (Baglin et al. 1992). A simple and effective vacuum load lock completes the assembly. Data acquisition is done with two ORTEC 918 multichannel buffer connected to an IBM-PC microcomputer. RUMP code for microcomputers (Doolittle, 1985; 1986; CGS, 1996) is used for spectra manipulation and analysis..

The PIXE station, depicted in Figure 2, is a home-made high vacuum chamber (15cm ID, 25cm high), that has been used since 1976 for PIXE analysis of air pollution at the Institute of Physics (Souza et al., 1976; Tabacniks, 1983). Initially installed on the 8MV Pelletron accelerator in the Nuclear Physics Department, and recently adapted to the new machine, the PIXE setup uses two Si(Li) KEVEX detectors with pulsed optical feedback for best resolution. One detector at 158° has a 7,1mm$^2$ crystal and resolution of 138 eV FWHM@MnK$\alpha$. A 50$\mu$m thick Be absorber is used to avoid backscattered protons from hitting the detector. This detector and geometry was optimized for low energy X-ray detection. The detector for high energy X-rays has a 25mm$^2$ crystal, resolution of 145eV FWHM@MnK$\alpha$, and is mounted at 90° to the target. A 550$\mu$m thick Mylar absorber is used to avoid low energy X-rays from being detected thus enhancing the high energy part of the spectrum on a clean background, without pile up events. Both detectors operate in vacuum. The use of two detectors respectively for low (1.3 to 5keV) and high (5 to 20keV) X-ray energies of the spectra, is intended to reduce data acquisition time while optimizing the X-ray yield and thus the detection limits. A computer controlled linear target holder for 18 samples completes the assembly. Inside the PIXE chamber are attachments to install surface barrier detectors that can be used for RBS measurements and/or beam monitoring when doing thick target analysis (Tabacniks, 1993; Martins and Tabacniks, 1993). The faraday cup accuracy for charge measurement was tested against the beam scattered in a thin gold foil into a surface barrier detector, with an agreement better than 1% (Martins, 1993). Data acquisition is done with an ORTEC 919 spectrum master ADC and multichannel, controlled by a Pentium type microcomputer. X-ray spectra analysis is done using AXIL software (V.Espen, et al.). PIXE analyses are usually run with 2.4MeV proton beam. Mass calibration is achieved analyzing homogeneous evaporated thin film standards (~50$\mu$g/cm$^2$) deposited on 2.5$\mu$m thick Mylar film, supplied by Micrommater, USA. Figure 3 shows the calibration curve for the São Paulo PIXE system. Figure 4 shows the result of 59 PIXE analysis of one multilayer target (Al/Ti/Zr/Mylar) used to monitor the calibration stability over a period of three months. Data were plotted in form of normalized residue to the average of the series according to $y = (x_i - \bar{x}) / \bar{x}$, where $x_i$ is each individual data point. Standard deviation was 10% for Al data 4.6%, for Ti and 3.3%. for Zr The higher deviation for Al and Ti may be respectively due to higher gradient of the X-ray ionization cross section in respect to energy.

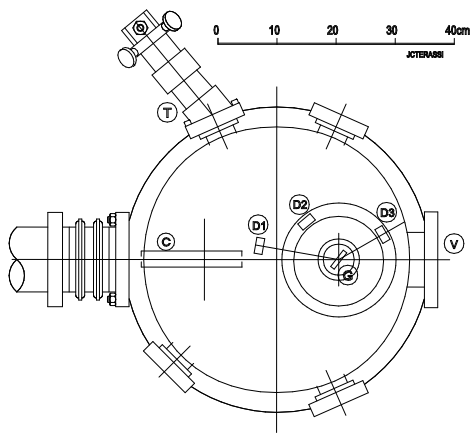

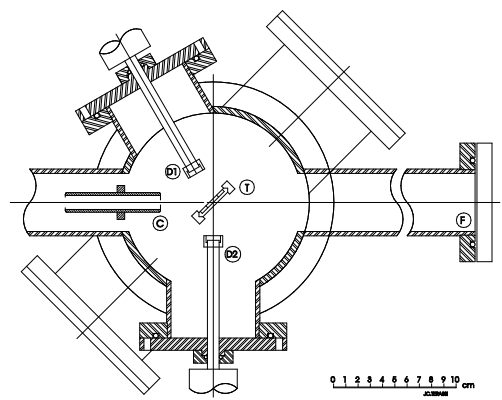

Figure 1. The RBS setup at LAMFI. (D1) fixed 170° SB detector, (D2) mobile SB detector, (D3) surface barrier for FRS analysis; (G) goniometer and target holder; (V) view port; (T) telescope for target positioning; (C) beam colimator.

Figure 2. PIXE setup at LAMFI. (C) beam collimator, (F) faraday cup, (T) target holder, (D1) low energy X-ray Si(Li) detector, (D2) high energy X-ray Si(Li) detector.

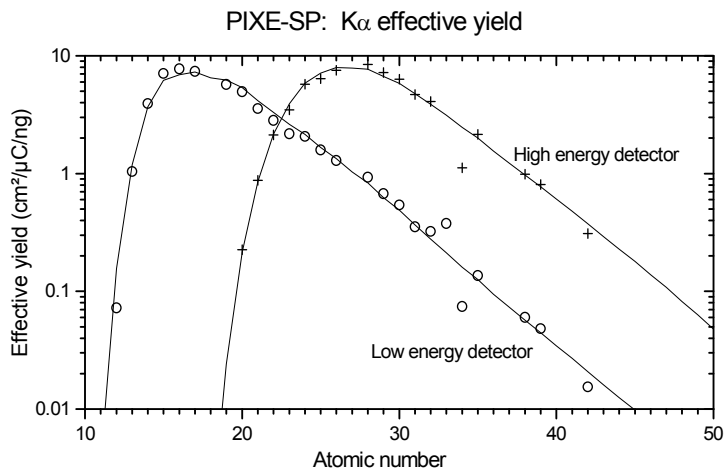

Figure 3. Calibration curves for the São Paulo PIXE system (adapted from Nascimento, 1997).